\documentclass[aps,prd,reprint,groupedaddress,amsmath,amssymb,showpacs,
floatfix]{revtex4-1}
\usepackage{graphicx}
\usepackage{dcolumn}
\usepackage{bm}
\begin{document}
\title{Light deflection due to a charged, rotating body}
\author{Sarani Chakraborty}
\email[]{sarani.chakraborty.phy@gmail.com}
\affiliation{Department of Physics, Assam University, Silchar-788011, Assam, India.}
\author{A. K. Sen}
\email[]{asokesen@yahoo.com}
\affiliation{Department of Physics, Assam University, Silchar-788011, Assam, India.}
\date{\today}
\begin{abstract}
According to GTR and subsequent developments in the field, it is known that there are three factors namely mass, rotation and charge that can influence the space-time geometry. Accordingly, we discuss the effect of space-time geometry of a charged, rotating body on the motion of the light ray. We obtained the expression for equatorial deflection of light due to such a body up to fourth order term. In our expression for deflection angle it is clear that charge can influence the path of light ray. We used the null geodesic approach of light ray  for our calculation. If we set the charge to zero our expression of bending angle gets reduced to the Kerr equatorial bending angle.If we set rotation to zero our expression reduces to Resinner-Nordstr$\ddot{o}$m deflection angle and if we set both charge and rotation to zero our expression reduces to Schwarzschild bending angle. However, we get non-zero bending angle for a hypothetical massless, rotating, charged body.
\end{abstract}

% insert suggested PACS numbers in braces on next line
% insert suggested keywords - APS authors don't need to do this
%\keywords{}
%\maketitle must follow title, authors, abstract, \pacs, and \keywords
\maketitle
% body of paper here - Use proper section commands
% References should be done using the \cite, \ref, and \label commands
\section{INTRODUCTION}
General relativity, a theory of gravitation was developed by Einstein as early as in 1907 and the final form was given in 1915. One of the very important consequences of general relativity is bending of light ray in presence of gravitational field. Three factors that can influence the path of light ray are mass distribution, charge and spin. The exact solution of Einstein's field equation for a static, spherical, uncharged body was found by Schwarzschild in 1915 [1], for an uncharged rotating body was found by Kerr in 1963 [2] and for a rotating charged body was found by Newman which is known as Kerr-Newman metric [3,4]. The solution for Einstein field equation for a static, charged,spherically symmetric body was obtained by Reissner and Nordstr$\ddot{o}$m independently, known as  Reissner-Nordstr$\ddot{o}$m solution[5,6]. Einstein himself calculated bending angle $(\alpha=4GM/Rc^{2})$ of light up to first order term, assuming null geodesic path of light in space-time manifold, a basic postulate for light in general relativity. Where G is the gravitational constant, M is the mass of the gravitating body, c is the speed of light in free space and R is the distance of the closest approach. The observational verification of his prediction was made in 1919 during the total solar eclipse [7].\\After Einstein, different authors calculated the light deflection angle up to second or higher order terms. Keeton et al [8] calculated the higher order terms of light deflection angle for a Schwarzschild mass. Iyer et al [9]calculated it for strong field and under weak field approximation their expression matches with that of Keeton et al [8].
\\Bending angle for Kerr mass in equatorial plane was calculated by Iyer et al.[10,11] using null geodesic of photon. According to their result, deflection produced in presence of a rotating black hole explicitly depends on direction of motion of the light. If light ray is moving in the direction of spin, deflection angle is higher than the zero rotation Schwarzschild field and if the light ray is moving in opposite to the rotation, bending angle is smaller than Schwarzschild field. Bozza [12] obtained the lensing formula, and calculated the relativistic image position for a light ray trajectory close to equatorial plane of a Kerr black hole. Aazami et al [13,14] calculated the two components of light bending angle, along the direction of equatorial plane and perpendicular to the equatorial plane of a Kerr black hole in quasi-equatorial regime. All the above mentioned calculations were done using the null geodesic of photon.
\\On the other hand some authors have used material medium approach where the gravitational effect on light ray was calculated by assuming some effective refractive index assigned to the medium through which light is propagating. Atkinson [15] used this method to study the trajectory of light ray near a very massive, static and spherically symmetric star. Fischback et al [16] calculated the second order contribution to gravitational deflection by a static mass using the same method. Sen[17] used this method to calculate the gravitational deflection of light without any weak field approximation. Balaz [18] used this method to calculate the change in the direction of polarization vector of electromagnetic wave passing close to a rotating body.
\\All above mentioned calculations were done for Schwarzschild and Kerr mass. Virbhadra et al [19] worked on Janis-Newman-Winicour (JNW) mass which is a charged, static mass and calculated the light deflection angle up to second order. Eiroa et al [20] worked on Reissner-Nordstr$\ddot{o}$m mass and calculated light deflection angle in both strong and weak deflection limit.
\\Hasse et al [21] worked on the lensig by Kerr-Newman mass and showed that infinite number of images formed by such body. Kraniotis [22] derived the analytic solutions of the lens equations in terms of Appell and Lauricella hypergeometric functions and the Weierstraß modular form for a Kerr-Newman mass. He used this formula to calculate the light deflection angle. But he did not apply any boundary condition (zero spin, zero charge, both spin and charge zero) to the expression of deflection angle [22, equation no. 87] to verify his result. In this paper we calculate exact expression for light deflection angle by a Kerr-Newman mass in equatorial plane which is a function of mass, spin and charge. We use null geodesic of photon approach to calculate the light deflection angle and verify our result through above mentioned boundary conditions.
\section{Kerr-Newman line element}
In GTR, Kerr-Newman line element represents the most generalised form of space-time curvature where all three factors (mass distribution, rotation, charge) have their contribution.
\\ Kerr-Newman line element is,
$$ds^{2}=(1-\frac{2mr-Q^{2}}{r^{2}+a^{2}\cos^{2}\vartheta})c^{2}dt^{2}-\frac{r^{2}+a^{2}\cos^{2}\vartheta}{r^{2}-2mr+a^{2}+Q^{2}}dr^{2}$$
 $$-(r^{2}+a^{2}\cos^{2}\vartheta)d\vartheta^{2}-(r^{2}+a^{2}$$
 $$+\frac{a^{2}\sin^{2}\vartheta(2mr-Q^{2})}{r^{2}+a^{2}\cos^{2}\vartheta})\sin^{2}\vartheta d\varphi^{2}$$
 $$+\frac{2a(2mr-Q^{2})\sin^{2}\vartheta}{r^{2}+a^{2}\cos^{2}\vartheta}dtd\varphi$$
 In equatorial plane $\vartheta=\frac{\pi}{2}$, so the modified line element is
 \begin{widetext}
 \begin{equation}
  ds^{2}=(1-\frac{2m}{r}+\frac{Q^{2}}{r^{2}})c^{2}dt^{2}+(\frac{4ma}{r}-\frac{2aQ^{2}}{r^{2}})cdtd\varphi-(\frac{r^{2}}{\Delta})dr^{2}
  -(r^{2}+a^{2}+\frac{2ma^{2}}{r}-\frac{Q^{2}a^{2}}{r^{2}})d\varphi^{2}
 \end{equation}
 \end{widetext}
 where,$\Delta=r^{2}-2mr+a^{2}+Q^{2}$, $m=\frac{GM}{c^{2}}$ and $a=\frac{J}{cM}$, further c, G, M and J are the velocity of light in free space, gravitational constant, mass and angular momentum of the gravitating body.
 Lagrangian (S) of such system [23, page 96] is given by:
 \begin{equation}
 2S=g_{\mu\nu}\frac{dx^{\mu}}{d\tau}\frac{dx^{\nu}}{d\tau}
 \end{equation}
 here $\tau$ represents the affine parameter. Thus,
 \begin{equation}
 2S=(1-\frac{2m}{r}+\frac{Q^{2}}{r^{2}})c^{2}\dot{t}^{2}+(\frac{4ma}{r}-\frac{2aQ^{2}}{r^{2}})c\dot{t}\dot{\varphi}$$
 $$-(\frac{r^{2}}{\Delta})\dot{r}^{2}-(r^{2}+a^{2}+\frac{2ma^{2}}{r}-\frac{Q^{2}a^{2}}{r^{2}})\dot{\varphi}^{2}
 \end{equation}
 where, dot (.) indicates differentiation with respect to $\tau$. Let $E$ and $L$ be the energy and angular momentum of the light ray along the direction of spin axis of the gravitating body, sign of $L$ decides the direction of motion of the light ray [10]. Then we can use equation (3) to obtain generalized momenta which are given as [23, page no. 327],
 $$p_{t}=\frac{\partial S}{\partial (c \dot{t})}=E$$ and $$-p_{\varphi}=-\frac{\partial S}{\partial \dot{\varphi}}=L$$

 That gives,
 \begin{equation}
 (1-\frac{2m}{r}+\frac{Q^{2}}{r^{2}})c\dot{t}+(\frac{2ma}{r}-\frac{aQ^{2}}{r^{2}})\dot{\varphi}=E
 \end{equation}
 and,
 \begin{equation}
 -(\frac{2ma}{r}-\frac{aQ^{2}}{r^{2}})c\dot{t}+(r^{2}+a^{2}+\frac{2ma^{2}}{r}-\frac{Q^{2}a^{2}}{r^{2}})\dot{\varphi}=L
 \end{equation}
 Solving the above two equations for $c\dot{t}$ and $\dot{\varphi}$ we get,
 \begin{equation}
 \dot{\varphi}=\frac{1}{\Delta}[(1-\frac{2m}{r}+\frac{Q^{2}}{r^{2}})L+(\frac{2ma}{r}-\frac{aQ^{2}}{r^{2}})E]
 \end{equation}
 And
 \begin{equation}
 c\dot{t}=\frac{1}{\Delta}[(r^{2}+a^{2}+\frac{2ma^{2}}{r}-\frac{Q^{2}a^{2}}{r^{2}})E-(\frac{2ma}{r}-\frac{aQ^{2}}{r^{2}})L]
 \end{equation}
 Again from equation (3),
 \begin{equation}
 -p_{r}=-\frac{\partial S}{\partial\dot{r}}=(\frac{r^{2}}{\Delta})\dot{r}
 \end{equation}
 Since for null geodesic,
$$ds^{2}=0$$
therefore,
$$g_{\mu\nu}dx^{\mu}dx^{\nu}=0$$
Or,
$$g_{\mu\nu}\frac{dx^{\mu}}{d\tau}\frac{dx^{\nu}}{d\tau}=0$$
But from equation (2),
$$g_{\mu\nu}\frac{dx^{\mu}}{d\tau}\frac{dx^{\nu}}{d\tau}=2S$$
$$2S=0$$
So the Lagrangian for null geodesic is zero, so equation (3) can be written as
 $$0=(1-\frac{2m}{r}+\frac{Q^{2}}{r^{2}})c^{2}\dot{t}^{2}+(\frac{4ma}{r}-\frac{2aQ^{2}}{r^{2}})c\dot{t}\dot{\varphi}$$
 $$-(\frac{r^{2}}{\Delta})\dot{r}^{2}-(r^{2}+a^{2}+\frac{2ma^{2}}{r}-\frac{Q^{2}a^{2}}{r^{2}})\dot{\varphi}^{2}$$
 Or,
\begin{equation}
 [(1-\frac{2m}{r}+\frac{Q^{2}}{r^{2}})c\dot{t}+(\frac{2ma}{r}-\frac{aQ^{2}}{r^{2}})]c\dot{t}-[(r^{2}+a^{2}+\frac{2ma^{2}}{r}$$
 $$-\frac{Q^{2}a^{2}}{r^{2}})\dot{\varphi}-(\frac{2ma}{r}-\frac{aQ^{2}}{r^{2}})]\dot{\varphi}-\frac{r^{2}}{\Delta}\dot{r}^{2}=0
\end{equation}
 Now from equation (4) and (5) we know, $$(1-\frac{2m}{r}+\frac{Q^{2}}{r^{2}})c\dot{t}+(\frac{2ma}{r}-\frac{aQ^{2}}{r^{2}})\dot{\varphi}=E$$ and $$(r^{2}+a^{2}+\frac{2ma^{2}}{r}-\frac{Q^{2}a^{2}}{r^{2}})\dot{\varphi}-(\frac{2ma}{r}-\frac{aQ^{2}}{r^{2}})c\dot{t}=L$$ So in equation (9), the quantity within first and second square bracket can be replaced by $E$ and $L$. So the new form of equation (9),
 \begin{equation}
 Ec\dot{t}-L\dot{\varphi}-\frac{r^{2}}{\Delta}\dot{r}^{2}=0
 \end{equation}
 Now taking the values of $\dot{\varphi}$ and $c\dot{t}$ from equations (6) and (7) and substituting them in equation (10) we have,
 \begin{widetext}
\begin{equation}
 \frac{r^{2}}{\Delta}\dot{r}^{2}=\frac{E}{\Delta}[(r^{2}+a^{2}+\frac{2ma^{2}}{r}-\frac{Q^{2}a^{2}}{r^{2}})E-(\frac{2ma}{r}-\frac{aQ^{2}}{r^{2}})L]$$
 $$-\frac{L}{\Delta}[(1-\frac{2m}{r}+\frac{Q^{2}}{r^{2}})L+(\frac{2ma}{r}-\frac{aQ^{2}}{r^{2}})E]
\end{equation}
\end{widetext}
or,
$$r^{2}\dot{r}^{2}=(r^{2}+a^{2}+\frac{2ma^{2}}{r}-\frac{Q^{2}a^{2}}{r^{2}})E^{2}-(\frac{2ma}{r}-\frac{aQ^{2}}{r^{2}})EL$$
$$-(1-\frac{2m}{r}+\frac{Q^{2}}{r^{2}})L^{2}-(\frac{2ma}{r}-\frac{aQ^{2}}{r^{2}})EL$$
or,
$$r^{2}\dot{r}^{2}=r^{2}E^{2}+(a^{2}E-2aEL+L^{2})(\frac{2m}{r}-\frac{Q^{2}}{r^{2}})-(L^{2}-a^{2}E^{2})$$
re-arranging we have,
$$r^{2}\dot{r}^{2}=r^{2}E^{2}+(aE-L)^{2}(\frac{2m}{r}-\frac{Q^{2}}{r^{2}})-(L^{2}-a^{2}E^{2})$$
or,
$$\dot{r}^{2}=E^{2}+\frac{1}{r^{4}}(aE-L)^{2}(2mr-Q^{2})-\frac{1}{r^{2}}(L^{2}-a^{2}E^{2})$$
It is possible to show that impact parameter is the ratio of $L$ and $E$ [23, page 328]. Following [10, equation no. 7], we can write the impact parameter $b_{s}=s(\frac{L}{E})=sb$, where $s=+1$ for prograde and $s=-1$ for retrograde orbit of light ray and $b$ is the positive magnitude of the impact parameter. But $b_{s}^{2}=(sb)^{2}=(\pm b)^{2}=b^{2}$, so we can drop $s$ for the even power of impact parameter. So the new form of the above equation is,
$$\dot{r}^{2}=L^{2}[\frac{1}{b^{2}}+\frac{1}{r^{4}}(\frac{a}{b_{s}}-1)^{2}(2mr-Q^{2})-\frac{1}{r^{2}}(1-\frac{a^{2}}{b^{2}})]$$
or,
\begin{equation}
 \dot{r}=L[\frac{1}{b^{2}}+\frac{1}{r^{4}}(\frac{a}{b_{s}}-1)^{2}(2mr-Q^{2})-\frac{1}{r^{2}}(1-\frac{a^{2}}{b^{2}})]^{\frac{1}{2}}   \end{equation}
 r obtains a local extremum for the closest approach $r_{o}$, we can write:
 $$\dot{r}|_{r=r_{0}}=0$$
 Thus, from equation (9),
 $$\frac{1}{b^{2}}+\frac{1}{r_{0}^{4}}(\frac{a}{b_{s}}-1)^{2}(2mr_{0}-Q^{2})-\frac{1}{r_{0}^{2}}(1-\frac{a^{2}}{b^{2}})=0$$
 or,
 \begin{equation}
 \frac{r_{0}^{2}}{b^{2}}=(1-\frac{a^{2}}{b^{2}})-(\frac{a}{b_{s}}-1)^{2}(\frac{2m}{r_{0}}-\frac{Q^{2}}{r_{0}^{2}})
 \end{equation}
\section{Light Deflection Angle}
The light deflection angle can be expressed as [24, page: 188],
\begin{equation}
 \alpha=2\int_{r_{o}}^{\infty}(\frac{d\varphi}{dr}).dr -\pi
\end{equation}
Now using equation (6) and (12)
\begin{widetext}
\begin{equation}
\frac{d\varphi}{dr}=\frac{(1-\frac{2m}{r}+\frac{Q^{2}}{r^{2}})+\frac{1}{b_{s}}(\frac{2ma}{r}-\frac{aQ^{2}}{r^{2}})}{\Delta\sqrt{\frac{1}{b^{2}}+\frac{1}{r^{4}}(\frac{a}{b_{s}}-1)^{2}(2mr-Q^{2})-\frac{1}{r^{2}}(1-\frac{a^{2}}{b^{2}})}}
\end{equation}
Using equation (15) in equation (14) we will get
\begin{equation}
 \alpha=2\int_{r_{o}}^{\infty}\frac{(1-\frac{2m}{r}+\frac{Q^{2}}{r^{2}})+\frac{1}{b_{s}}(\frac{2ma}{r}-\frac{aQ^{2}}{r^{2}})}{\Delta\sqrt{\frac{1}{b^{2}}+\frac{1}{r^{4}}(\frac{a}{b_{s}}-1)^{2}(2mr-Q^{2})-\frac{1}{r^{2}}(1-\frac{a^{2}}{b^{2}})}}.dr-\pi
\end{equation}
\end{widetext}
Let us introduce a new variable $x=\frac{r_{o}}{r}$. So,
$$dx=-\frac{r_{0}dr}{r^{2}}$$
or,
$$\frac{dx}{r_{0}}=-\frac{dr}{r^{2}}$$
the limits will change as' when $r\longrightarrow \infty$, then $x\longrightarrow 0$ and when $r\longrightarrow r_{0}$, then $x\longrightarrow 1$. Using this in above equation we have,
\begin{widetext}
\begin{equation}
 \alpha=2\int_{0}^{1}\frac{(1-\frac{2mx}{r_{0}}+\frac{Q^{2}x^{2}}{r_{0}^{2}})+\frac{1}{b_{s}}(\frac{2max}{r_{0}}-\frac{aQ^{2}x^{2}}{r_{0}^{2}})}{(1-\frac{2mx}{r_{0}}+\frac{Q^{2}x^{2}}{r_{0}^{2}}+\frac{a^{2}x^{2}}{r_{0}^{2}})r_{0}\sqrt{\frac{1}{b^{2}}+\frac{x^{4}}{r_{0}^{4}}(\frac{a}{b_{s}}-1)^{2}(2mr-Q^{2})-\frac{x^{2}}{r_{0}^{2}}(1-\frac{a^{2}}{b^{2}})}}.dx-\pi
\end{equation}
\end{widetext}
Let us substitute $h=\frac{m}{r_{o}}$ and $n^{2}=\frac{Q^{2}}{r_{0}^{2}}$ and $\hat{a}=\frac{a}{m}$. So the new format of equation (17) is
\begin{widetext}
\begin{equation}
 \alpha=2\int_{0}^{1}\frac{(1-2hx+x^{2}n^{2})+\frac{1}{b_{s}}(2hax-an^{2}x^{2})}{(1-2hx+x^{2}n^{2}+\hat{a}^{2}h^{2}x^{2})\sqrt{\frac{r_{0}^{2}}{b^{2}}+x^{2}(\frac{a}{b_{s}}-1)^{2}(2hx-x^{2}n^{2})-x^{2}(1-\frac{a^{2}}{b^{2}})}}.dx-\pi \end{equation}
 \end{widetext}
Let us put the expression of $\frac{r_{0}^{2}}{b^{2}}=(1-\frac{a^{2}}{b^{2}})-(\frac{a}{b_{s}}-1)^{2}(\frac{2m}{r_{0}}-\frac{Q^{2}}{r_{0}^{2}})=(1-\frac{a^{2}}{b^{2}})-(\frac{a}{b_{s}}-1)^{2}(2h-n^{2})$ from equation (13) in equation (18), we get,
 \begin{widetext}
 $$\alpha=2\int_{0}^{1}\frac{(1-2hx+x^{2}n^{2})+\frac{1}{b_{s}}(2hax-an^{2}x^{2})}{(1-2hx+x^{2}n^{2}+\hat{a}^{2}h^{2}x^{2})\sqrt{(1-\frac{a^{2}}{b^{2}})-(\frac{a}{b_{s}}-1)^{2}(2h-n^{2})+x^{2}(\frac{a}{b_{s}}-1)^{2}(2hx-x^{2}n^{2})-x^{2}(1-\frac{a^{2}}{b^{2}})}}.dx-\pi$$
 rearranging above equation we have,
 \begin{equation}
  \alpha=2\int_{0}^{1}\frac{1-2hx(1-\frac{a}{b_{s}})+x^{2}n^{2}(1-\frac{a}{b_{s}})}{(1-2hx+x^{2}n^{2}+\hat{a}^{2}h^{2}x^{2})\sqrt{(1-\frac{a^{2}}{b^{2}})(1-x^{2})-(1-\frac{a}{b_{s}})^{2}2h(1-x^{3})+n^{2}(1-\frac{a}{b_{s}})^{2}(1-x^{4})}}.dx-\pi
 \end{equation}
 \end{widetext}
 Now following [13,14], let us consider $G=1-(\frac{a}{b})^{2}=1-\hat{a}^{2}(\frac{m}{b})^{2}$ and $F=1-(\frac{a}{b_{s}})=1-s\hat{a}\frac{m}{b}$. Thus for zero rotation $(\hat{a}=0)$, $F=G=1$. Now the new form of equation (19) using $F$ and $G$ is,
 \begin{widetext}
 \begin{equation}
 \alpha=2\int_{0}^{1}\frac{1-2Fhx+Fn^{2}x^{2}}{(1-2hx+x^{2}n^{2}+\hat{a}^{2}h^{2}x^{2})\sqrt{G(1-x^{2})-2F^{2}h(1-x^{3})+F^{2}n^{2}(1-x^{4})}}.dx-\pi
 \end{equation}
 or,
 \begin{equation}
 \alpha=2\int_{0}^{1}\frac{1-2Fhx+Fn^{2}x^{2}}{(1-2hx+x^{2}n^{2}+\hat{a}^{2}h^{2}x^{2})\sqrt{G}\sqrt{1-x^{2}}\sqrt{1-\frac{2F^{2}h(1-x^{3})}{G(1-x^{3})}+\frac{F^{2}n^{2}(1+x^{2})}{G}}}.dx-\pi
 \end{equation}
 \end{widetext}
 Or,
 \begin{widetext}
 $$\alpha=2\int_{0}^{1}\frac{1-2Fhx+Fn^{2}x^{2}}{(1-2hx+x^{2}n^{2}+\hat{a}^{2}h^{2}x^{2})\sqrt{G}\sqrt{1-x^{2}}\sqrt{1-\frac{2F^{2}h(1-x^{3})}{G(1-x^{3})}}\sqrt{1+\frac{F^{2}n^{2}(1+x^{2})}{G}(1-\frac{2F^{2}h(1-x^{3})}{G(1-x^{3})})^{-1}}}.dx-\pi$$
 \end{widetext}
 Now rearranging the above equation we can write.
\begin{widetext}
$$\alpha=2\int_{0}^{1}\frac{dx}{\sqrt{G}\sqrt{1-x^{2}}}[1-2Fhx+Fn^{2}x^{2}][1-2hx+x^{2}n^{2}+\hat{a}^{2}h^{2}x^{2}]^{-1}[1-\frac{2F^{2}h(1-x^{3})}{G(1-x^{3})}]^{-\frac{1}{2}}$$
$$[1+\frac{F^{2}n^{2}(1+x^{2})}{G}(1-\frac{2F^{2}h(1-x^{3})}{G(1-x^{3})})^{-1}]^{-\frac{1}{2}}-\pi$$
or,
$$\alpha=2\int_{0}^{1}\frac{dx}{\sqrt{G}\sqrt{1-x^{2}}}[1-2Fhx+Fn^{2}x^{2}][1-2hx+\hat{a}^{2}h^{2}x^{2}]^{-1}[1+x^{2}n^{2}(1-2hx+\hat{a}^{2}h^{2}x^{2})^{-1}]^{-1}[1-\frac{2F^{2}h(1-x^{3})}{G(1-x^{3})}]^{-\frac{1}{2}}$$
$$[1+\frac{F^{2}n^{2}(1+x^{2})}{G}(1-\frac{2F^{2}h(1-x^{3})}{G(1-x^{3})})^{-1}]^{-\frac{1}{2}}-\pi$$
\end{widetext}
For weak deflection limit, following [13,14], $Q,m\ll r_{o}$, i.e we can say that, $h,n\ll 1$. So the above equation can be expanded in Taylor series in terms of both $h$ and $n$. Here we calculate up to fourth order terms only,
\begin{widetext}
$$\alpha=2\int_{0}^{1}\frac{dx}{\sqrt{G}\sqrt{1-x^{2}}}[1-2Fhx+Fn^{2}x^{2}][1+2hx+x^{2}h^{2}(4-\hat{a}^{2})+x^{3}h^{3}(8-4\hat{a}^{2})$$
$$+x^{4}h^{4}(\hat{a}^{4}-12\hat{a}^{2}+16)][1-x^{2}n^{2}\{1+2hx+x^{2}h^{2}(4-\hat{a}^{2})\}+x^{4}n^{4}]$$
$$[1+\frac{F^{2}h(1-x^{3})}{G(1-x^{2})}+\frac{3F^{4}h^{2}(1-x^{3})^{2}}{2G^{2}(1-x^{2})^{2}}$$
$$+\frac{5F^{6}h^{3}(1-x^{3})^{3}}{2G^{3}(1-x^{2})^{3}}+\frac{35F^{8}h^{4}(1-x^{3})^{4}}{8G^{4}(1-x^{2})^{4}}][1$$
$$-\frac{F^{2}n^{2(1+x^{2})}}{2G}\{1+\frac{2F^{2}h(1-x^{3})}{G(1-x^{2})}+\frac{4F^{4}h^{2}(1-x^{3})^{2}}{G^{2}(1-x^{2})^{2}}\}$$
$$+\frac{3F^{4}n^{4}}{G}(1+x^{2})^{2}]-\pi$$
\end{widetext}
Multiplying term by term taking up to fourth order of both $h$ and $n$ we get,
\begin{widetext}
$$\alpha=2\int_{0}^{1}\frac{dx}{\sqrt{G}\sqrt{1-x^{2}}}[1+h\{2x(1-F)+\frac{F^{2}}{G}(\frac{1-x^{3}}{1-x^{2}})-4x^{3}n^{2}(1-F)-\frac{F^{2}n^{2}(1-F)}{G}(\frac{1-x^{3}}{1-x^{2}})$$
$$-\frac{F^{2}n^{2}x(1+x^{2})}{G}(1-F)-\frac{3F^{4}n^{2}(1+x^{2})}{2G^{2}}(\frac{1-x^{3}}{1-x^{2}})\}$$
$$+h^{2}\{-n^{2}x^{4}(4-\hat{a}^{2})(1-F)-4x^{4}n^{2}(1-F)+x^{2}(4-\hat{a}^{2}-4F)-x^{4}n^{2}(4-\hat{a}^{2}-4F)-\frac{2F^{2}n^{2}x^{3}(1-F)}{G}(\frac{1-x^{3}}{1-x^{2}})+\frac{2F^{2}x(1-F)}{G}(\frac{1-x^{3}}{1-x^{2}})$$
$$-\frac{2F^{2}x^{3}n^{2}}{G}(\frac{1-x^{3}}{1-x^{2}})+\frac{3F^{4}}{2G^{2}}(\frac{1-x^{3}}{1-x^{2}})^{2}-\frac{3F^{4}x^{2}n^{2}(1-F)}{2G}(\frac{1-x^{3}}{1-x^{2}})^{2}$$
$$-\frac{2F^{4}n^{2}x(1+x^{2})(1-F)}{G}(\frac{1-x^{3}}{1-x^{2}})-\frac{F^{2}n^{2}x^{2}(1+x^{2})}{2G}(4-\hat{a}^{2}-4F)-\frac{F^{4}n^{2}x(1+x^{2})(1+x^{2})(1-F)}{G^{2}}(\frac{1-x^{3}}{1-x^{2}})$$
$$-\frac{15F^{6}n^{2}(1+x^{2})}{4G^{3}}(\frac{1-x^{3}}{1-x^{2}})^{2}\}+h^{3}\{x^{3}(-4\hat{a}^{2}+8+2F\hat{a}^{2}-8F)+\frac{F^{2}x^{2}}{G}(\frac{1-x^{3}}{1-x^{2}})(4-\hat{a}^{2}-4F)$$
$$+\frac{3F^{4}x(1-F)}{G^{2}}(\frac{1-x^{3}}{1-x^{2}})^{2}+\frac{5F^{6}}{2G^{3}}(\frac{1-x^{3}}{1-x^{2}})^{3}\}+h^{4}\{x^{4}(\hat{a}^{4}-12\hat{a}^{2}+16+8\hat{a}^{2}F-16F)+\frac{F^{2}x^{3}}{G}(\frac{1-x^{3}}{1-x^{2}})(8-4\hat{a}^{2}+2F\hat{a}^{2}-8F)$$
$$+\frac{3F^{4}x^{2}}{2G^{2}}(\frac{1-x^{3}}{1-x^{2}})^{2}(4-\hat{a}^{2}-4F)+\frac{5xF^{6}(1-F)}{G^{3}}(\frac{1-x^{3}}{1-x^{2}})^{3}+\frac{35F^{8}}{8G^{4}}(\frac{1-x^{3}}{1-x^{2}})^{4}\}+n^{2}\{-x^{2}(1-F)-\frac{F^{2}}{2G}(1+x^{2})\}$$
$$+n^{4}\{x^{4}(1-F)+\frac{F^{2}x^{2}}{2G}(1+x^{2})(1-F)+\frac{3F^{4}}{8G^{2}}(1+x^{2})^{2}\}]-\pi$$
\end{widetext}
Integrating term by term we get,
\begin{widetext}
\begin{equation}
\alpha=(\frac{1}{\sqrt{G}}-1)\pi+4h[\frac{F^{2}+G-FG}{G^{\frac{3}{2}}}-(\frac{7}{2}-\frac{3\pi}{8})\frac{F^{4}n^{2}}{G^{\frac{5}{2}}}-\frac{(1-F)}{\sqrt{G}}\{\frac{4n^{2}}{3}-(\frac{41}{12}-\frac{\pi}{4})\frac{F^{2}n^{2}}{G}\}]$$
$$+h^{2}[-4\{\frac{F^{2}}{G^{\frac{5}{2}}}(F^{2}+G-FG)\}+\frac{15\pi}{4}\frac{1}{15G^{\frac{5}{2}}}\{15F^{4}-4G(F-1)(3F^{2}+2G^{2})-2G^{2}\hat{a}^{2}\}-(-50+\frac{825\pi}{32})\frac{F^{6}n^{2}}{G^{3}}+(1-F)n^{2}\{-\frac{3\pi}{8}(4-\hat{a}^{2})$$
$$-\frac{3\pi}{2}-(-16+\frac{15\pi}{2})\frac{F^{2}}{G}-\frac{F^{4}}{G^{2}}(\frac{105\pi}{16}-16)-(-18+\frac{81\pi}{8})\frac{F^{4}}{G^{2}}\}-\frac{7\pi}{16}\frac{F^{2}n^{2}}{G}(-\hat{a}^{2}+4-4F)]$$
$$+h^{3}[\frac{122}{3}\frac{1}{61G\frac{7}{2}}\{61F^{6}-G(F-1)(45F^{4}+32F^{2}G+16G^{2})-4G^{2}\hat{a}^{2}(2F^{2}+2G-FG)\}-\frac{15\pi}{2}\frac{F^{2}}{G}\frac{1}{15G^{\frac{5}{2}}}\{15F^{4}-4G(F-1)(3F^{2}+2G^{2})$$
$$-2\hat{a}^{2}G^{2}\}]+h^{4}[-130\frac{F^{2}}{65G^{\frac{9}{2}}}\{65F^{6}-49(F-1)F^{4}G-8F^{2}G^{2}(-4+\hat{a}^{2}+4F)+4(4+\hat{a}^{2}(F-2)-4F)G^{3}\}+\frac{3465\pi}{64}\frac{1}{1155G^{\frac{9}{2}}}\{1155F^{8}$$
$$-840(F-1)F^{6}G-140F^{4}(-4+\hat{a}^{2}+4F)G^{2}+80(4+\hat{a}^{2}(F-2)-4F)F^{2}G^{3}+8(16-12\hat{a}^{2}+\hat{a}^{4}+8(\hat{a}^{2}-2)F)G^{4}\}]+n^{2}[-\frac{\pi}{2}(1-F)-\frac{3\pi F^{2}}{4G}]$$
$$+n^{4}[\frac{3\pi}{8}(1-F)+\frac{7\pi}{16}\frac{F^{2}(1-F)}{G}+\frac{57\pi}{64}]
\end{equation}
\end{widetext}
The above equation (22) is the general expression for deflection of light due to rotating charged sphere in equatorial plane. This expression is a function of mass, rotation and charge. We verify our result with some boundary conditions in the following section of the paper.
\\If we set charge equal to zero in equation (22), we will get,
\begin{widetext}
\begin{equation}
 \alpha=(\frac{1}{\sqrt{G}}-1)\pi+4h[\frac{F^{2}+G-FG}{G^{\frac{3}{2}}}]+h^{2}[-4\{\frac{F^{2}}{G^{\frac{5}{2}}}(F^{2}+G-FG)\}+\frac{15\pi}{4}\frac{1}{15G^{\frac{5}{2}}}\{15F^{4}-4G(F-1)(3F^{2}+2G^{2})-2G^{2}\hat{a}^{2}\}]$$
$$+h^{3}[\frac{122}{3}\frac{1}{61G\frac{7}{2}}\{61F^{6}-G(F-1)(45F^{4}+32F^{2}G+16G^{2})-4G^{2}\hat{a}^{2}(2F^{2}+2G-FG)\}-\frac{15\pi}{2}\frac{F^{2}}{G}\frac{1}{15G^{\frac{5}{2}}}\{15F^{4}-4G(F-1)(3F^{2}+2G^{2})$$
$$-2\hat{a}^{2}G^{2}\}]+h^{4}[-130\frac{F^{2}}{65G^{\frac{9}{2}}}\{65F^{6}-49(F-1)F^{4}G-8F^{2}G^{2}(-4+\hat{a}^{2}+4F)+4(4+\hat{a}^{2}(F-2)-4F)G^{3}\}+\frac{3465\pi}{64}\frac{1}{1155G^{\frac{9}{2}}}\{1155F^{8}$$
$$-840(F-1)F^{6}G-140F^{4}(-4+\hat{a}^{2}+4F)G^{2}+80(4+\hat{a}^{2}(F-2)-4F)F^{2}G^{3}+8(16-12\hat{a}^{2}+\hat{a}^{4}+8(\hat{a}^{2}-2)F)G^{4}\}]
\end{equation}
\end{widetext}
This is the expression of deflection for light by Kerr mass obtained by Aazami et al [14, equation no. (B17)].
\\If we set rotation equal to zero with non-zero charge in equation (22), we get,
\begin{widetext}
\begin{equation}
 \alpha=4h+(-4+\frac{15\pi}{4})h^{2}+(\frac{122}{3}-\frac{15\pi}{2})h^{3}+(-130+\frac{3465\pi}{64})h^{4}-\frac{3\pi}{4}n^{2}
+\frac{57\pi}{64}n^{4}-(14-\frac{3\pi}{2})n^{2}h-(-50+\frac{825\pi}{32})n^{2}h^{2}
\end{equation}
\end{widetext}
This is the expression for deflection by a charged, non-rotating body. This expression, up to second order contribution of charge i.e $n^{2}$ was obtained by Eiroa et al [20, equation no. 55].
\\If we set both charge and rotation equal to zero we get,
\begin{equation}
\alpha=4h+(-4+\frac{15\pi}{4})h^{2}+(\frac{122}{3}-\frac{15\pi}{2})h^{3}+(-130+\frac{3465\pi}{64})h^{4}
\end{equation}
This is the well known expression of light deflection angle due to Schwarzschild mass given by Keeton et al [8, equation no. 23].
\\If we set mass and rotation equal to zero in equation (22), we will get,
\begin{equation}
 \alpha=(\frac{1}{\sqrt{G}}-1)\pi+n^{2}[-\frac{\pi}{2}(1-F)-\frac{3\pi F^{2}}{4G}]$$
$$+n^{4}[\frac{3\pi}{8}(1-F)+\frac{7\pi}{16}\frac{F^{2}(1-F)}{G}+\frac{57\pi}{64}]
\end{equation}
This is the amount of deflection of light ray occurring only due to charge. A hypothetical massless rotating body can influence the curvature of space time.
\section{conclusion}
From above study we may conclude the following:\\
1. Charge has a noticeable effect on the path of the light ray. When compared with Kerr expression for bending, we find that there are some extra terms in the expression for deflection which occur due to the presence of charge. If we set the charge equal to zero,deflection angle will reduce to that of Kerr deflection angle. If we set both charge and rotation equal to zero, deflection angle will reduce to that of Schwarzschild deflection angle.\\
2. If we set mass equal to zero, our bending angle does not reduce to zero, i.e. charge itself can influence the path of light ray. So we can say that only charge without mass can influence the curvature of space-time.\\
3. For zero mass, contribution of rotation parameter $a$ is not zero as $a$ occurs not only with mass but also with charge in both Kerr-Newman line element and expression for light deflection angle in equation (22). So we can say that such kind of massless, charged, rotating body it self can produce frame dragging, gravitational redshift and time dilation without any contribution of mass. However, the concept of a massless, charged, rotating body is purely hypothetical.
\begin{acknowledgments}
Some of the calculations reported in this paper were done using the online version of Mathematica. S. Chakraborty deeply acknowledge S. Roy, Assam University, Silchar for useful discussions and Dr. A. Deshamukhya, Head Dept. of Physica, Assam University, Silchar, India for her encouragement to carry out this work.
\end{acknowledgments}

% Create the reference section using BibTeX:
%\bibliography{basename of .bib file}

\end{document}